\documentclass[reprint,aps,pra]{revtex4-1}

\usepackage{graphicx}
\usepackage{dcolumn}
\usepackage{bm}
\usepackage{amsmath}
\usepackage{amsthm}
\usepackage{amssymb}
\usepackage{float}
\usepackage{hyperref}
\usepackage{color}
\usepackage{epstopdf}
\usepackage{cleveref}
\usepackage[svgnames]{xcolor}
\hypersetup{hidelinks,colorlinks=true,allcolors=DarkBlue}

\begin{document}

\preprint{APS/123-QED}

\title
{
Towards events recognition in a distributed fiber-optic sensor system: \\
Kolmogorov--Zurbenko filtering
}
\author{A.K. Fedorov$^{1}$}
\author{M.N. Anufriev$^{1}$}
\author{A.A. Zhirnov$^{1}$}
\author{E.T. Nesterov$^{1}$}
\author{D.E. Namiot$^{2}$}
\author{V.E. Karasik$^{1}$}
\author{A.B. Pnev$^{1}$}
\affiliation
{
\mbox{$^{1}$Bauman Moscow State Technical University, 2nd Baumanskaya St. 5, Moscow 105005, Russia}
\mbox{$^{2}$Lomonosov Moscow State University, Vorob'evy gory 1, Moscow 119992, Russia}
}

\date{\today}

\begin{abstract}
The paper is about de-noising procedures aimed on events recognition 
in signals from a distributed fiber-optic vibration sensor system based on the phase-sensitive optical time-domain reflectometry.
We report experimental results on recognition of several classes of events in a seismic background. 
A de-noising procedure uses the framework of the time-series analysis and Kolmogorov--Zurbenko filtering. 
We demonstrate that this approach allows revealing signatures of several classes of events.
\end{abstract}
                              
\maketitle

\section{Introduction}
	
Real-time monitoring systems with the use of distributed fiber-optic vibration sensor systems based on the phase-sensitive optical time-domain reflectometry technique 
\cite{Barnoski,Costa,Barnoski2,Healey,Healey2,Healey3}. 
have a fascinating prospective for applications. 
Examples include control for access on protected areas ({\it e.g.}, securing national borders), oil and gas pipelines, communications lines, and structural health monitoring \cite{Raj,Cau,Bao}. 

The core of such system is phase-sensitive optical time-domain reflectometry technique, which has sufficiently high sensitivity and spatial resolution 
\cite{Tai,Takada,Juskaitis,Kulchin,Koyamada,Martins,Alekseev,Shi,Duan}. 
A main feature of this type of reflectometry is a sufficiently large coherence length of the employed optical pulse. 
Signals reflected from centers of the Rayleigh backscattering exhibit the coherent summation of their complex wave amplitudes. 

On the one hand, monitoring systems based on the phase-sensitive optical time-domain reflectometry are sensitive enough to register sufficiently small fluctuations \cite{Healey3}. 
On the other hand, this means that an algorithm, which allows revealing the nature of fluctuations, should supplement such systems. 
Indeed, the crucial problem here is to reveal: are these fluctuations caused by natural changes of background or by any kind of activates? 
In other words, a highly non-trivial problem of de-noising comes to the fore.

To make a decision about the nature of fluctuations one can continuously analyze signals from the system in time or frequency domains. 
Due to a sufficiently complex structure of signals from the monitoring system this problem is rather challenging. 
It has been extensively studied during last decade (see \cite{Shi,Li,Lyons} and reference therein). 
However, at this moment there is no universal solution for events recognition problem for vibration sensor systems based on the phase-sensitive optical time-domain reflectometry technique.

Mathematically speaking, signals from distributed fiber-optic vibration sensor systems are time series \cite{Namiot}. 
Then for their analysis various types of time series analysis can be applied. 
In this paper, we present experimental results on an application of the Kolmogorov-Zurbenko filtering \cite{Zurbenko,Zurbenko2,Zurbenko3,Zurbenko4,Zurbenko5}
for signal de-noising in a fiber-optic distributed vibration sensor system based on the phase-sensitive optical time-domain reflectometry. 
In the considered case, the main goal of the system is a control for access on protected areas. 
Consequently, we are confronted with the problem of events recognition ({\it e.g.}, human passage, human group passage, or car travel) in a seismic background. 

The paper is organized as follows. 
In Sec. \ref{sec:setup}, we describe our setup for collecting experimental data and parameters of the fiber-optic distributed vibration sensor system. 
In Sec. \ref{sec:signals}, we describe the basic de-noising procedure based on the Kolmogorov-Zurbenko filtering. 
In Sec. \ref{sec:basic}, we present experimental results on application of the Kolmogorov-Zurbenko de-noising procedure to measured signals. 
In Sec. \ref{sec:conclusion}, we give our conclusion.

\begin{figure}[t]
\center
\includegraphics[width=0.95\columnwidth]{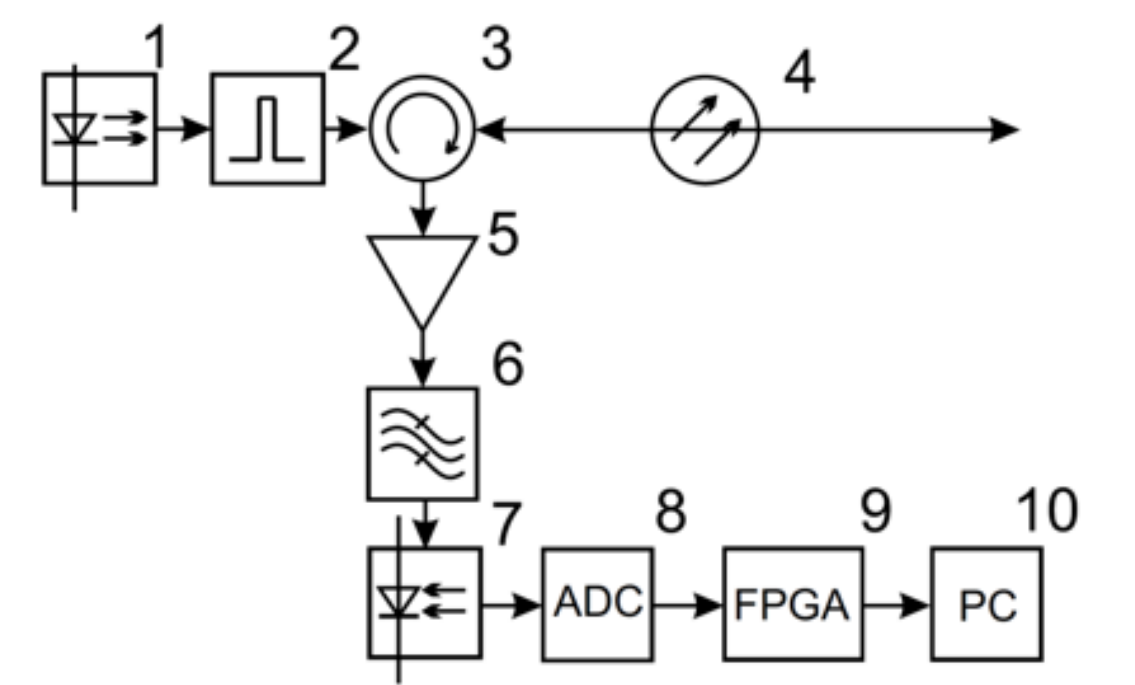}
\vskip -2mm
\caption
{
	Fiber-optic distributed sensor system based on the phase-sensitive optical time-domain reflectometry: setup for collecting of experimental data. 
	The sensing element of the system is the standard single mode fiber (fiber-optic cable).
}
\label{fig:setup}
\end{figure}

\section{Setup for collecting experimental data}\label{sec:setup}

The fiber-optic distributed vibration sensor system for collecting of data is located on the Bauman Moscow State Technical University polygon in Moscow Region. 
The setup is presented on Fig. \ref{fig:setup}: 
1 is the primary light source (laser), 
2 is the acousto-optic modulator, 
3 is the circulator, 
4 is the fiber-optic sensor, 
5 is the preamplifier, 
6 is the optical filter, 
7 is the detector, 
8 is the ADC converter, 
9 is the programmable logic device, 
10 is the computer. 

Probe signal has the wavelength 1550 nm, probe pulse of duration 200 ns, and the signal from the semiconductor laser of power 300-500 mW is launched into the standard optical fiber. 
Probe signal has ultra-narrow line width, which is less than 1 MHz. In our experiments, length of the optical fiber cable l is approximately 50 km.

The idea of its work can be presented as follows. In case of a vibration impact, the intensity of backscattering light changes according to the level of the impact. 
Circulator is used for launching of the probe signal into the optical fiber and the backscattering signal to a detector. 
Signals in the system are sum of all scattered signals during time of pulse with taking into account their phases. 

\section{Signals from the system: events recognition}\label{sec:signals}

\begin{figure}[h]
\center
\includegraphics[width=1\columnwidth]{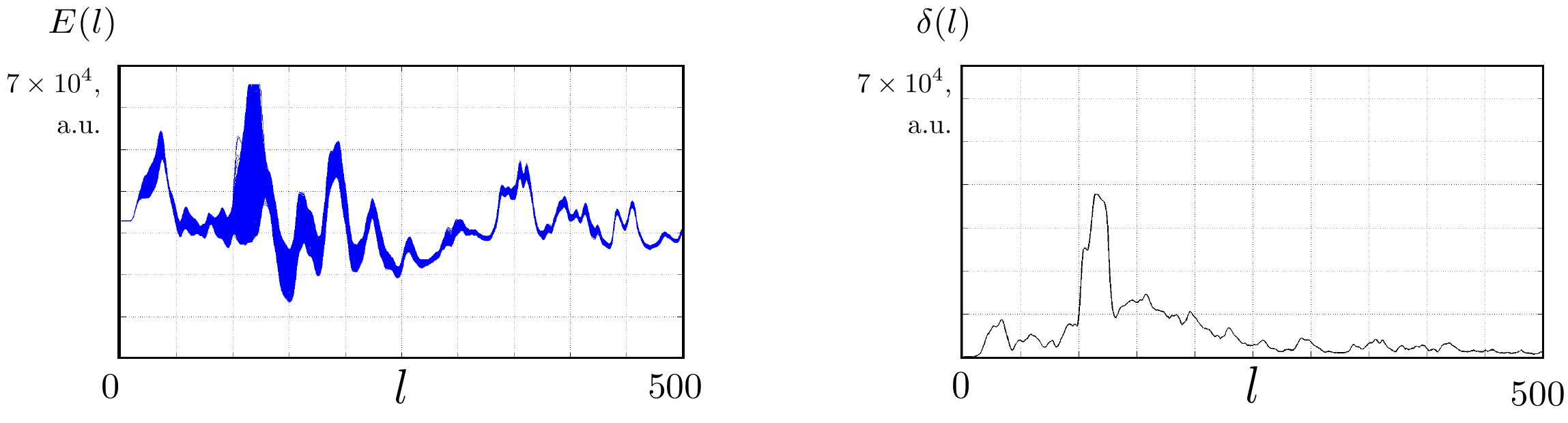}
\vskip -4mm
\caption
{
	An overlap of typical signals measured by the system on the region of the cable with length 0.5 km during one second (left) and the maximum deviation (right): 
	a signature of an event.
}
\label{fig:overlap}
\end{figure}

A typical result of the measurement using the setup is presented in Fig. \ref{fig:overlap}a. 
This figure presents an overlap between signals measured by our setup (Sec. \ref{sec:setup}) on the region of the cable with length 0.5 km during one second. 

As it was mentioned above, the crucial challenge is to recognize any kind of deliberate activity in these signals. 
It is clear that this problem can be essentially divided on two related sub-problems. 
The first part is to register the event in a (seismic) background. 
It is seen from Fig. \ref{fig:overlap}b that an event can be registered by the system via measurement of the difference between signals in neighboring moments of time. 
From such a procedure, one can find, {\it e.g.}, position of the event in the cable. 
However, due to natural fluctuation of the background such simple procedure leads to a sufficiently large number of false positives. 

The second problem is to classify registered event. In this work, we are guided by the following basic classification of events: 
(i) single event, which is localized in space and in time; 
(ii) single event, which is delocalized in space and localized in time; 
(iii) single event, which is localized in space and delocalized in time; 
(iv) single event, which is delocalized in space and delocalized in time.

For our setup an important part is using of preliminary tests. 
These test consist of collecting of a large number of experimental data corresponding to background and typical types of activates. 
Preliminary test have been organized on the polygon at similar experimental conditions (importantly, the same climate conditions). 
From preliminary tests, we use a sufficiently large number of experimentally measured signals of type 
(i)-(iv) to obtain minimum, maximum, and average characteristics values (characteristic time scales of events and characteristic length scales) as well as parameters of the background. 

\begin{figure}[h]
\center
\includegraphics[width=0.975\columnwidth]{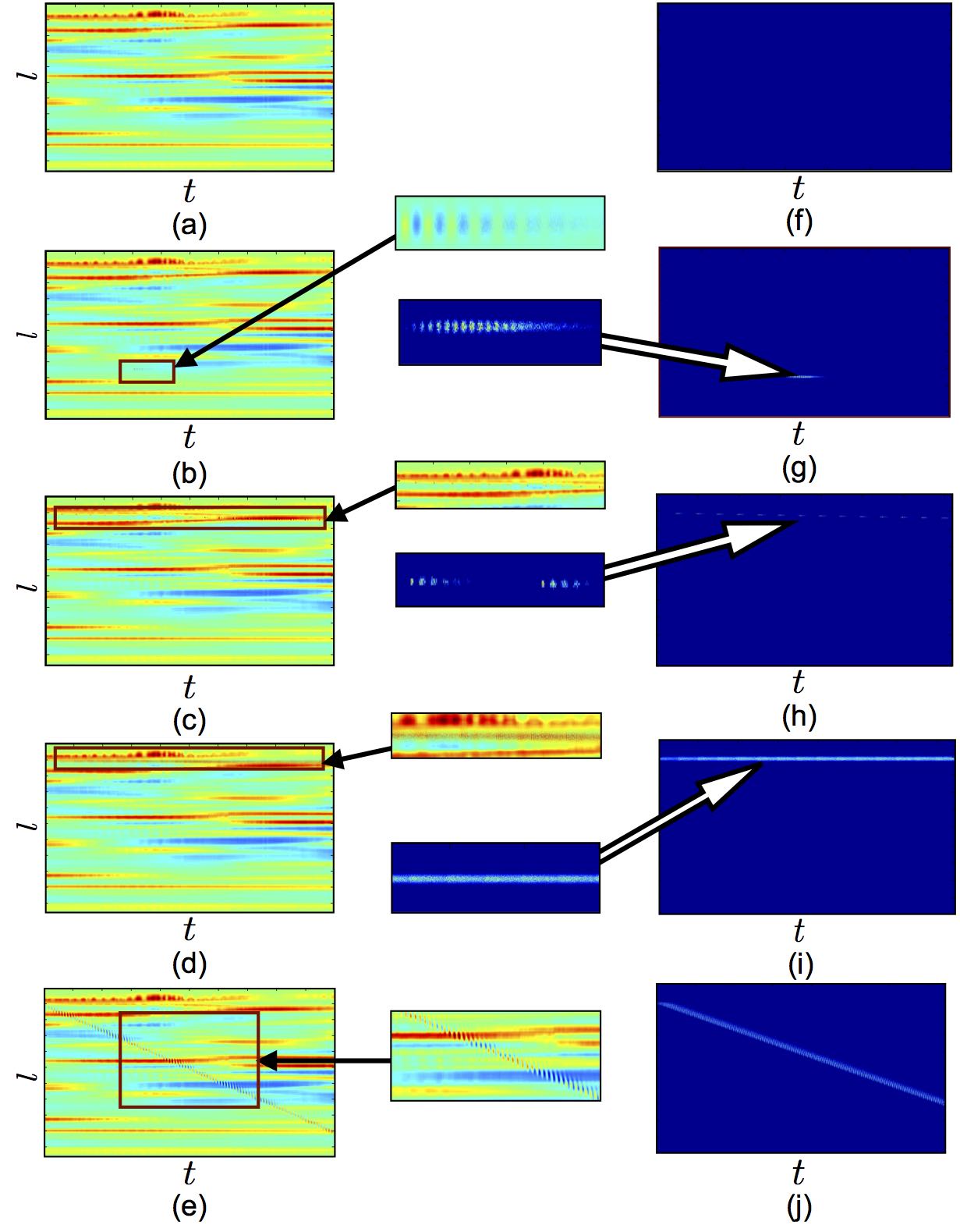}
\vskip -4mm
\caption
{
	Measured signals \cite{data}: 
	(a) background signals experimentally measured by the system; 
	(b) sequence of space localized short-term events with additive noise form background; 
	(c) sequence of space delocalized short-term events with additive noise form background; 
	 space localized long-term event with additive noise form background; 
	(e) space delocalized long-term event with additive noise form background; 
	results of the de-noising procedure (f)-(j) for the signals (a)-(e).
}
\label{fig:results}
\end{figure}  

The background noise is the waterfall form, {\it i.e.}, intensity as a function of time (t) and position in the fiber-optic cable (l), is presented in Fig. \ref{fig:results}a. 
A sequence of space-time localized events is presented in Fig. \ref{fig:results}b. 
Fig. \ref{fig:results}c shows a sequence of delocalized in space and localized in time events. 
In Fig. \ref{fig:results}d, one can see a sequence of localized in space and delocalized in time events. 
Fig. \ref{fig:results}e shows a sequence of space-time delocalized events.

\section{Basic algorithm}\label{sec:basic}

Signals from a distributed fiber-optic vibration sensor systems are time series. 
Therefore, for their analysis various types of time series analysis can be employed. 
One of the possible solutions is to use statistical apparatus of the time series analysis. 

For instants, an event can be detected in the background in the difference in the value of signals between time moments $t=j$ and $t=j+1$. 
Toward this end, we suggest to implement the spatial low-pass filter of the Kolmogorov--Zurbenko type \cite{Zurbenko,Zurbenko2,Zurbenko3,Zurbenko4,Zurbenko5}. 
Similar approaches to de-noising procedure based on moving average calculation have been recently discussed \cite{MA,MA2}.

\begin{figure}[h]
\center
\includegraphics[width=0.95\columnwidth]{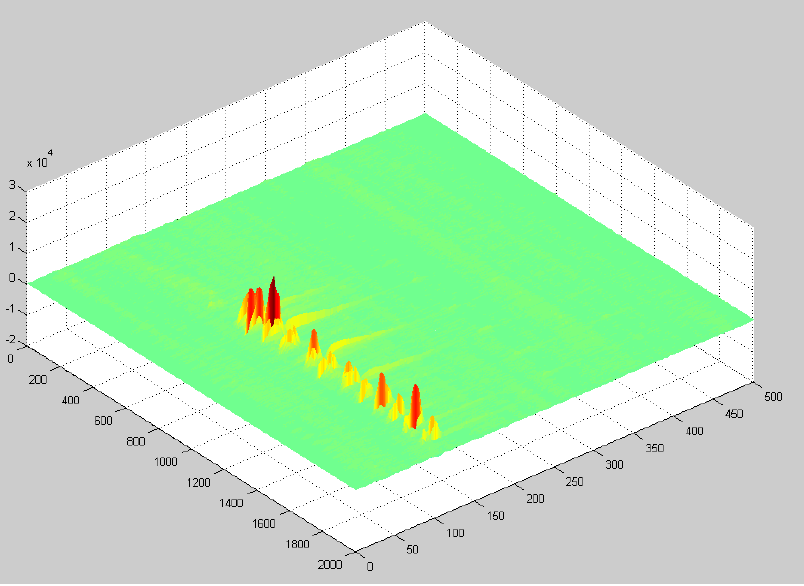}
\vskip -2mm
\caption
{
	3D visualization of the signal measured form the system after the de-noising procedure based on the basic algorithm (1)-(4) and classified 
	by the system as sequence of space and time localized events, which can correspond to human passage along the cable. 
}
\label{fig:results3D}
\end{figure}   

In general, the Kolmogorov--Zurbenko filter is a series of iterations of the moving average filter. 
The first iteration of Kolmogorov--Zurbenko filter is an application of the moving average filter over a given process.
\begin{equation}\label{eq:KZF}
	\mathbf{KZ}\left[X(t,l)\right]_{m,k=1}=\sum_{s=-(m-1)/2}^{(m-1)/2}{X(t+s,l)\,\frac{1}{m}}
\end{equation}
Here, $X(l,t)$ is the signal (real-valued time series), $m$ is the time window, and k is the filter order \cite{Zurbenko}. 
The iteration process of a simple operation of moving average is very computationally efficient. 
The Kolmogorov--Zurbenko filter has been employed in investigations of climate fluctuations and seminal studies of the turbulence problem
\cite{Zurbenko,Zurbenko2,Zurbenko3,Zurbenko4,Zurbenko5}. 

Using the de-noising procedure based on the application of the Kolmogorov--Zurbenko filter (\ref{eq:KZF}) to signals, 
the sensor system can detect an event in the background. 
Indeed, if the moving average deviation of the signal $X(l,t)$ 
\begin{equation}\label{eq:delta}
	\delta_j^{\rm MA}=\left|{X(t,l)-\mathbf{KZ}\left[X(t,l)\right]}\right|
\end{equation}
exceeds the critical value, then a potential event is detected. 
The key question for implementation of such an approach is about determination of the critical value (\ref{eq:delta}). 

We suggest use an adaptive critical value calculation in a rather straightforward way. 
Let us consider the deference between two signals with a time window $j$ in the following form:
\begin{equation}\label{eq:deltaa}
	\delta_j(\Delta{l})=\left|{X(t+j,l)-X(t,l)}\right|.
\end{equation}
By integrating this difference over the cable length
\begin{equation}\label{eq:deltaaint}
	\int{\delta_j(\Delta{l})dl}=\delta^{*}
\end{equation}
one can obtain the critical value, which characterizes the integrated difference between two signals from the system in neighboring countdowns of time. 
On practice, we obtain critical values on the bases of preliminary tests of the system (see Sec. \ref{sec:signals}).  

This critical value is continuously measured by the monitoring system during preliminary tests. 
Finally, from the de-noising procedure with taking into account the critical value, we obtain the de-noised signal as follows:
\begin{equation}\label{eq:result}
	S(l,t)=\left\{\begin{array}{ll}
	\mathbf{KZ}\left[X(t,l)\right],	&\mbox{if $\delta_j^{\rm MA}>\delta^{*}$},\\
	0,	&\mbox{otherwise}.
	\end{array}\right.
\end{equation}

We optimize the de-noising procedure with respect to parameters of the Kolmogorov--Zurbenko filter: 
size of the time window and order of the filter. 
This allows obtaining the maximal level of recognized events.  

The results of the application of the de-noising procedure (\ref{eq:KZF})-(\ref{eq:deltaaint}) 
for experimentally measured background signals with embedded events \cite{data} of type (i)-(iv) are presented in Fig. \ref{fig:results}. 
In spite of simplicity of used de-noising procedure, events can be clearly detected. 
Furthermore, patterns of different classes of events can be reliably revealed (See Fig. \ref{fig:results3D}).
 
\section{Results and conclusion}\label{sec:conclusion} 

Controllability of a workflow is the most important requirement for its efficient implementation.  
Then a problem of a design of real-time monitoring systems for nondestructive testing has a paramount importance for technological processes in a broad class of applications in industry. 
Mainly, the goal of a monitoring system is to measure and analyze specific characteristics of a process, which allows to make a decision about the correctness of its functioning. 

In this work, we have investigated the basic algorithmic solution for recognition of events in seismic backgrounds. 
Performance of the suggested algorithmic solution has been demonstrated on experimentally measured signals.


\begin{thebibliography}{9}
\bibitem{data}
Data for experimental evidence of the suggested algorithm have been prepared as follows.
Simulated events, which parameters correspond to parameters of measured in experiment events, 
are embedded in the experimentally measured background. 
This procedure has been implemented to illustrate revealing of different types of events at explicitly the same experimental conditions including the same level of self-noise.
Nontrivial effects related to changing of these conditions rely an additional type of filtering procedure. 
These problems will be considered in another place.
\end{thebibliography}
\end{document}